# A Single-Crystalline, Epitaxial SrTiO$_3$ Thin-Film Transistor

Kosuke Uchida[1], Akira Yoshikawa[1], Kunihito Koumoto[1], Takeharu Kato[2], Yuichi Ikuhara[2,3], and Hiromichi Ohta[1,4,a]

[1]*Graduate School of Engineering, Nagoya University, Chikusa, Nagoya 464-8603, Japan*

[2]*Japan Fine Ceramics Center, Mutsuno, Atsuta, Nagoya 456-8587, Japan*

[3]*Institute of Engineering Innovation, The University of Tokyo, Bunkyo, Tokyo 113-8656, Japan*

[4]*PRESTO, Japan Science and Technology Agency, Honcho, Kawaguchi 332-0012, Japan*

[a]Correspondence should be addressed H.O. (h-ohta@apchem.nagoya-u.ac.jp)

PACS: 81.15.Fg Pulsed laser ablation deposition, 73.61.Le Other inorganic semiconductors, 85.30.Tv Field effect devices, 77.55.D- High-permittivity gate dielectric films

**Abstract**

We report herein fabrication and characterization of a thin-film transistor (TFT) using single-crystalline, epitaxial $SrTiO_3$ film, which was grown by a pulsed laser deposition technique followed by the thermal annealing treatment in an oxygen atmosphere. Although TFTs on the polycrystalline epitaxial $SrTiO_3$ films (as-deposited) exhibited poor transistor characteristics, the annealed single-crystalline $SrTiO_3$ TFT exhibits transistor characteristics comparable with those of bulk single-crystal $SrTiO_3$ FET: an on/off current ratio $>10^5$, sub-threshold swing ~2.1 Vdecade$^{-1}$, and field-effect mobility ~0.8 $cm^2V^{-1}s^{-1}$. This demonstrates the effectiveness of the appropriate thermal annealing treatment of epitaxial $SrTiO_3$ films.

Strontium titanate (SrTiO$_3$, cubic perovskite, *Pm*3*m*, lattice constant $a$ = 3.905 Å) is known as a band insulator with a wide bandgap of ~3.2 eV. SrTiO$_3$ has attracted growing attention for the next generation of *oxide electronics*[1] because it exhibits several unique properties. Charge carrier concentration of SrTiO$_3$ can be easily varied from insulating to metallic ($n_{3D}$ ~10$^{21}$ cm$^{-3}$) by appropriate substitution doping, such as Nb$^{5+}$ (Ti$^{4+}$ site) or La$^{3+}$ (Sr$^{2+}$ site).[2,3] In addition, SrTiO$_3$ exhibits extremely high Hall mobility of >10$^4$ cm$^2$V$^{-1}$s$^{-1}$ at low temperatures.[4] High-quality single crystals of SrTiO$_3$, which is commercially available, are widely used in the heteroepitaxial film growth of several perovskite oxides, such as high-$T_c$ cuprates and manganates. The recent finding of high-density two-dimensional electron gas (2DEG) by Ohtomo and Hwang,[5] confined within an extremely thin layer at the LaAlO$_3$/SrTiO$_3$ heterointerface, further accelerates the motivation for the realization of SrTiO$_3$-based electronic devices.

One of the most important devices to prove the viability of SrTiO$_3$ is the field-effect transistor (FET). A number of SrTiO$_3$-based FETs have been reported to date, using bulk single crystals of SrTiO$_3$.[6–11] Very recently, Ueno and co-workers observed a superconducting transition ($T_c$ ~0.4 K) of electrostatically accumulated 2D electron channel (sheet charge concentration $n_{2D}$ =1−10 × 10$^{13}$ cm$^{-2}$) in a SrTiO$_3$ single crystal with an electric double layer gating technique.[12] They modulated the mean depth of carrier distribution (channel thickness) down from 16 to 3 nm, which corresponds to only 7−40 unit cells of SrTiO$_3$. This shows that a SrTiO$_3$-based thin-film transistor (TFT) is appropriate for practical 2D electron channel applications, since channel thickness of the TFT can be reduced further. However, SrTiO$_3$-based TFTs have not yet been reported.

In order to develop a SrTiO$_3$ TFT, we first tried to fabricate several top-gate TFTs by using ~60-nm thick epitaxial SrTiO$_3$ films, which were grown by pulsed laser deposition (PLD, KrF excimer laser, ~0.5 Jcm$^{-1}$pulse$^{-1}$, 20 ns, 5 Hz) on the (001) face of [(LaAlO$_3$)$_{0.3}$(Sr$_2$AlTaO$_6$)$_{0.7}$] (LSAT) substrates at 900ºC in an oxygen pressure of 5 × 10$^{-4}$ Pa.

We used amorphous 12CaO·7Al$_2$O$_3$ (*a*-C12A7, permittivity $\varepsilon_r$ = 12) as the gate insulator because single crystal SrTiO$_3$-FETs with such gates exhibit nearly ideal transistor characteristics (Ref. 11).

The resultant TFTs, however, exhibited normally-ON-type behavior with poor transistor characteristics. The on-off current ratio and sub-threshold swing S-factor were <10$^2$ and ~20 Vdecade$^{-1}$, respectively, which were worse than those of single-crystal SrTiO$_3$ FET,[11] most likely due to the high density of oxygen vacancies and/or charge traps in the SrTiO$_3$ film (data not shown). Furthermore, the surface of the films was composed of fine grains (~30 nm in diameter), although the stepped-and-terraced LSAT surface was also observed in atomic force microscope (AFM) images [Fig. 1(a)], indicating that layer-by-layer growth has occurred. In summary, the as-deposited SrTiO$_3$ films were oxygen-deficient, polycrystalline and epitaxial.

To overcome these problems, the as-deposited SrTiO$_3$ films were annealed at 900ºC in an oxygen pressure of ~1 Pa. We expected the polycrystalline film to be converted into a single crystal during thermal annealing via solid-phase diffusion. We obtained single-crystalline, epitaxial SrTiO$_3$ films with stepped-and-terraced surfaces. The resultant TFT exhibited characteristics comparable with those of a single-crystal SrTiO$_3$ FET: on/off current ratio greater than 10$^5$, S-factor ~2.1 Vdecade$^{-1}$, and field-effect mobility $\mu_{FE}$ ~0.8 cm$^2$V$^{-1}$s$^{-1}$. Here we report the fabrication and characterization of a SrTiO$_3$ TFT using a single-crystalline, epitaxial film of SrTiO$_3$.

First, polycrystalline, epitaxial SrTiO$_3$ films (thickness: ~60 nm) were grown on the (001) face of LSAT substrates by PLD as described above. After the film growth, pure O$_2$ gas (1 Pa) was introduced into the PLD chamber to fill the oxygen deficiency of the SrTiO$_3$ film. The film was then annealed at 900ºC for 30 min in an oxygen atmosphere and then cooled to room temperature. Figure 1(b) shows a topographic AFM image of the annealed film surface. The image shows atomically flat terraces and steps, which correspond to unit cell height of SrTiO$_3$ (0.3905 nm), without any grain boundaries. Crystallographic orientation and thickness of the films were evaluated by high-resolution X-ray diffraction (HRXRD, ATX-G, Rigaku Co.) using monochromated Cu K$\alpha_1$ radiation. Only the intense diffraction peak of 002 SrTiO$_3$ is observed, together with 002 LSAT in the out of plane Bragg diffraction pattern [Fig. 1(c)]. Pendellösung fringes, which indicate that the thickness of SrTiO$_3$ is 60 nm, are clearly seen around the 002 SrTiO$_3$ [inset of Fig. 1(c)]. From these results, we concluded that the polycrystalline epitaxial as-deposited film was converted into a single crystal by this thermal annealing.

We subsequently fabricated a top-gate TFT using single crystalline SrTiO$_3$ epitaxial film as shown in Fig. 2. First, 20-nm-thick, metallic Ti films were deposited for use as the source and drain electrodes. The deposition was performed through a stencil mask by electron beam (EB) evaporation (base pressure $\sim 10^{-4}$ Pa, no substrate heating). Second, 160-nm-thick *a*-C12A7 film was deposited through a stencil mask by PLD ($\sim 3$ Jcm$^{-2}$pulse$^{-1}$, oxygen pressure $\sim 0.1$ Pa) using dense polycrystalline C12A7 ceramic as a target. Finally, a gate electrode – a 20-nm-thick metallic Ti film – was deposited through a stencil mask by EB evaporation. The resultant TFT was annealed at 200ºC in air atmosphere to reduce the oxygen defects generated during the *a*-C12A7 deposition.

Figure 3 shows the cross sectional high-resolution transmission electron microscope (HRTEM, TOPCON EM-002B with an accelerating voltage of 200 kV) image of the Ti/*a*-C12A7/SrTiO$_3$/LSAT interfacial region. We clearly observe a multilayer structure. The thicknesses of Ti, *a*-C12A7 and SrTiO$_3$ are 20, 160 and 60 nm, respectively. The heterointerface between *a*-C12A7/SrTiO$_3$ is abrupt. While *a*-C12A7 is featureless, the crystal structure of the SrTiO$_3$ film is clearly observable. A nano-beam diffraction patterns corresponding to *a*-C12A7 indicates that *a*-C12A7 is a glass film.

Transistor characteristics of the resultant SrTiO$_3$ TFT were measured with a semiconductor device analyzer (B1500A, Agilent Technologies) at room temperature. The channel width (*W*) and the channel length (*L*) of the TFT were 400 and 200 μm, respectively. Figure 4 shows typical (a) transfer and (b) output characteristics of the resultant TFT. Drain current ($I_d$) of the FET increased as the gate voltage ($V_g$) increased, hence the channel was $n$-type, and electron carriers were accumulated by positive $V_g$ [Fig. 4(a)]. We observed a rather large threshold gate voltage ($V_{th}$) of +6.5 V, obtained from a linear fit of an $I_d^{0.5}$–$V_g$ plot [inset of (a)], which corresponds to an electron trapping state density of ~5 × 10$^{12}$ cm$^{-2}$. We also observed a clear pinch-off and current saturation in $I_d$ [Fig. 4(b)], indicating that the operation of this TFT conformed to standard FET theory. The on/off current ratio, S-factor and $V_{th}$ were >10$^5$, ~2.1 Vdecade$^{-1}$ and +6.5 V, respectively. We calculated the sheet charge concentration ($n_{xx}$) and the field-effect mobility ($\mu_{FE}$) of the SrTiO$_3$-TFTs. The $n_{xx}$ values were obtained from $n_{xx} = C_i(V_g - V_{th})$, where $C_i$ was the capacitance per unit area (67 nFcm$^{-2}$). The $\mu_{FE}$ values were obtained from $\mu_{FE} = g_m[(W/L)C_i \cdot V_d]^{-1}$, where $g_m$ was transconductance $\partial I_d/\partial V_g$. The maximum $\mu_{FE}$ for this TFT was ~0.8 cm$^2$V$^{-1}$s$^{-1}$.

These values of on/off current ratio, S-factor, $V_{th}$ and $\mu_{FE}$ are comparable to those of a single-crystal SrTiO$_3$ FET. The values for the as-deposited polycrystalline SrTiO$_3$-based TFT were significantly worse, demonstrating the effectiveness of appropriate thermal annealing treatment of epitaxial SrTiO$_3$ films.

In summary, we have fabricated the first thin-film transistor (TFT) which uses a single-crystalline epitaxial SrTiO$_3$ film. We found that polycrystalline epitaxial SrTiO$_3$ films can be converted into single crystals by thermal annealing. The resultant TFT exhibits transistor characteristics comparable to those of bulk single crystal SrTiO$_3$ FET: on/off current ratio >10$^5$, sub-threshold swing ~2.1 Vdecade$^{-1}$, and field-effect mobility ~0.8 cm$^2$V$^{-1}$s$^{-1}$. Appropriate thermal annealing treatment of epitaxial SrTiO$_3$ films in an oxygen atmosphere is necessary to reduce the number of oxygen vacancies and create single-crystal structures.

A part of this work was financially supported by MEXT (Nano Materials Science for Atomic-scale Modification, 20047007).

**FIG. 1 (Color online)** Topographic AFM images of SrTiO$_3$ epitaxial films [(a) as-deposited, (b) annealed in an oxygen atmosphere of 1 Pa]. (c) Out-of-plane XRD pattern of the SrTiO$_3$ epitaxial film. The inset shows Pendellösung fringes around the 002 diffraction peak of SrTiO$_3$.

**FIG. 2 (Color online)** The schematic device structure of a SrTiO$_3$ TFT. Ti films (20-nm-thick) are used as the source, drain and gate electrodes. A 160-nm-thick *a*-C12A7 film serves as the gate insulator. Channel length (*L*) and channel width (*W*) are 200 and 400 μm, respectively.

**FIG. 3 (Color online)** (a) Cross-sectional high-resolution transmission electron microscope image of the 160-nm-thick *a*-C12A7/SrTiO$_3$/LSAT heterointerface. (b) The magnified image around the *a*-C12A7/SrTiO$_3$ heterointerface. (c) Nano-beam diffraction pattern of *a*-C12A7 layer (upper) and selected area electron diffraction pattern of SrTiO$_3$ layer (lower).

**FIG. 4 (Color online)** (a) Typical transfer and (b) output characteristics of a TFT with a single-crystalline SrTiO$_3$ epitaxial film active layer, obtained by thermal annealing in an oxygen atmosphere. The inset of (a) shows $I_d^{0.5}$−$V_g$ plot of this TFT.

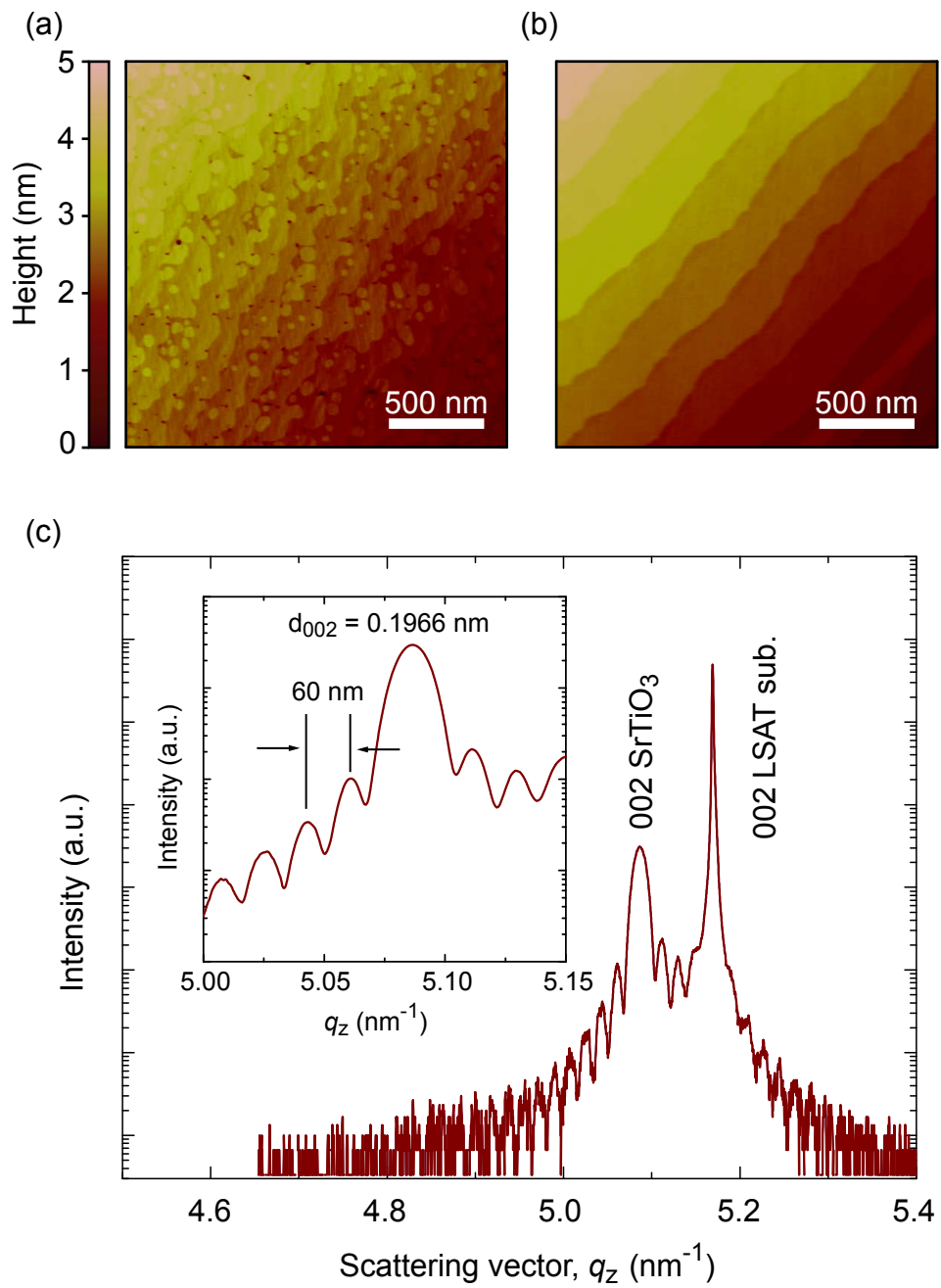

Fig. 1

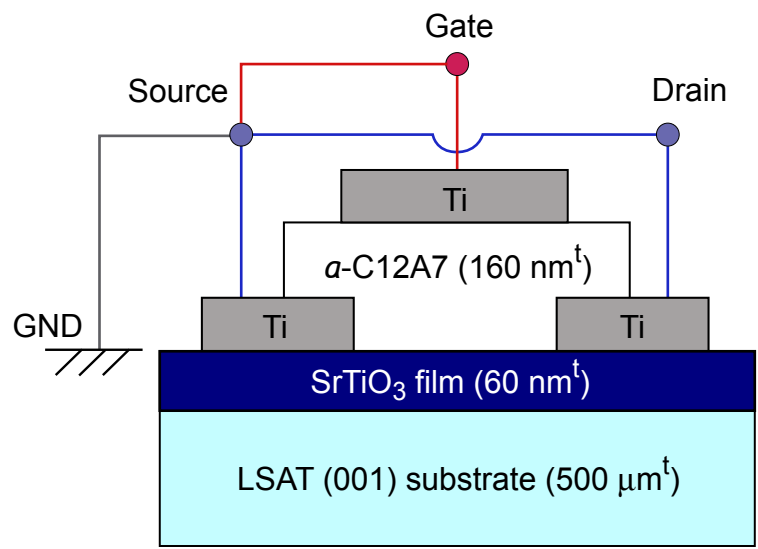

Fig. 2

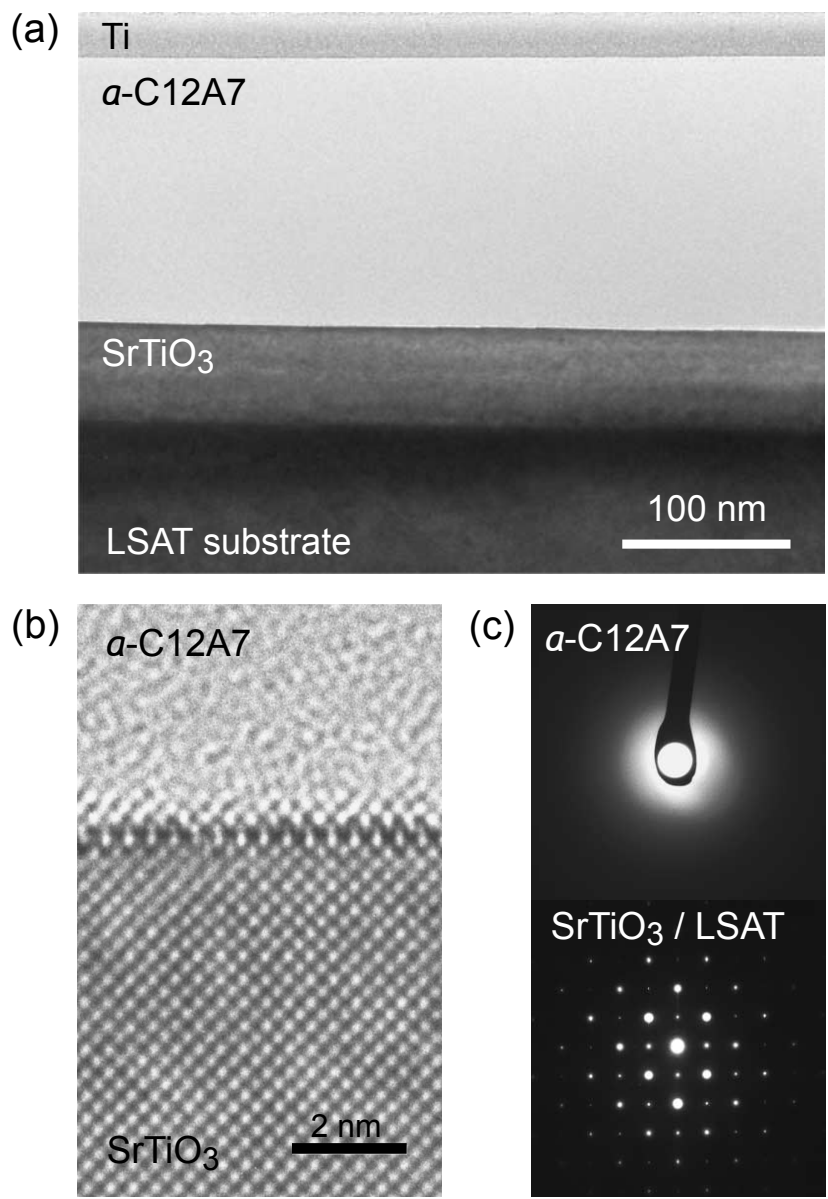

Fig. 3

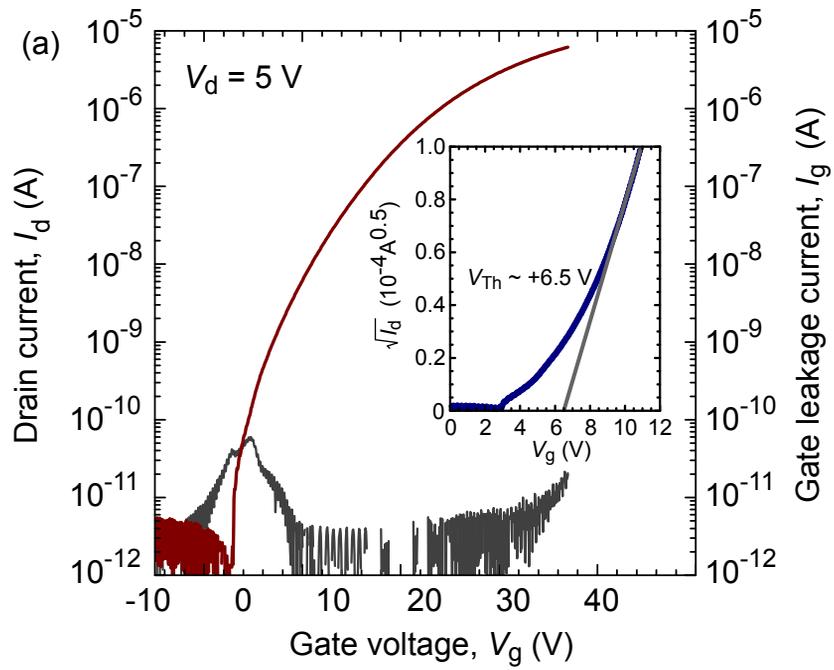

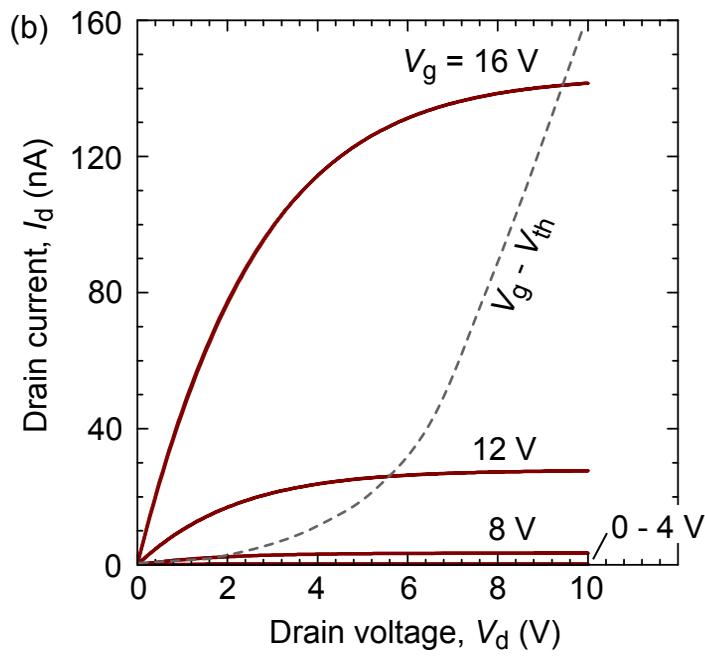

Fig. 4